\documentclass[twocolumn,amsmath,amssymb,floatfix,pra,aps]{revtex4}
\usepackage{graphicx}
\begin{document}

\title{A Bose-Einstein Condensate in a Uniform Light-induced Vector Potential}

\author{Y.-J.~Lin}
\author{R.~L.~Compton}
\author{A.~R.~Perry}
\author{W.~D.~Phillips}
\author{J.~V.~Porto}
\author{I.~B.~Spielman}
%\altaffiliation[To whom correspondence should be addressed: ]{@nist.gov}
\affiliation{National Institute of Standards and Technology,
Gaithersburg, MD 20899}

\date{\today}

\begin{abstract}
We use a two-photon dressing field to create an effective vector
gauge potential for Bose-condensed $^{87}$Rb atoms in the $F=1$
hyperfine ground state. The dressed states in this Raman field are
spin and momentum superpositions, and we adiabatically load the
atoms into the lowest energy dressed state. The effective
Hamiltonian of these neutral atoms is like that of charged particles
in a uniform magnetic vector potential, whose magnitude is set by
the strength and detuning of Raman coupling. The spin and momentum
decomposition of the dressed states reveals the strength of the
effective vector potential, and our measurements agree
quantitatively with a simple single-particle model. While the
uniform effective vector potential described here corresponds to
zero magnetic field, our technique can be extended to non-uniform
vector potentials, giving non-zero effective magnetic fields.

\end{abstract}

\maketitle

Ultracold atoms are an appealing system for the study of many-body
correlated states relevant to condensed matter physics. These widely
tunable systems in nearly disorder-free potentials have already
realized one-dimensional Tonks-Girardeau gases
\cite{paredes04Kinoshita04}, and the superfluid to Mott-insulator
transition of a Bose-Einstein condensate (BEC) in an optical lattice
\cite{SFMIref}. The implementation of these simple iconic condensed
matter systems paves the way for more interesting systems with
exotic correlations and excitations, as in the fractional quantum
Hall effect (FQHE) of a two-dimensional electron system in a strong
magnetic field \cite{tsui82laughlin83}. Simulating such a system
with neutral atoms requires an effective Lorentz force, which is
associated with a vector gauge potential. Current experimental
approaches involve rotation of trapped BECs
\cite{Aboshaeer01bretin04,schweikhard04}, where low field effects,
such as the formation of an Abrikosov vortex lattice, have been
observed. For technical reasons, this approach is limited to modest
effective fields, too small for FQHE
physics~\cite{schweikhard04,bloch08}. Most recent proposals to
create significantly larger effective magnetic fields without
rotation
\cite{jaksch03,sorensen05hafezi07,juzeliunas0406,zhu06,Higbie02}
involve optical coupling between internal atomic states, where the
atoms are dressed in a spatially dependent manner. The effective
magnetic field in such light-induced gauge potentials can be
understood as a consequence of changing into a spatially varying
basis of internal states. This is similar to the effective magnetic
field in a rotating BEC arising from changing into the frame
rotating with the BEC.

\begin{figure*}
\begin{center}
\includegraphics[width=6.6in]{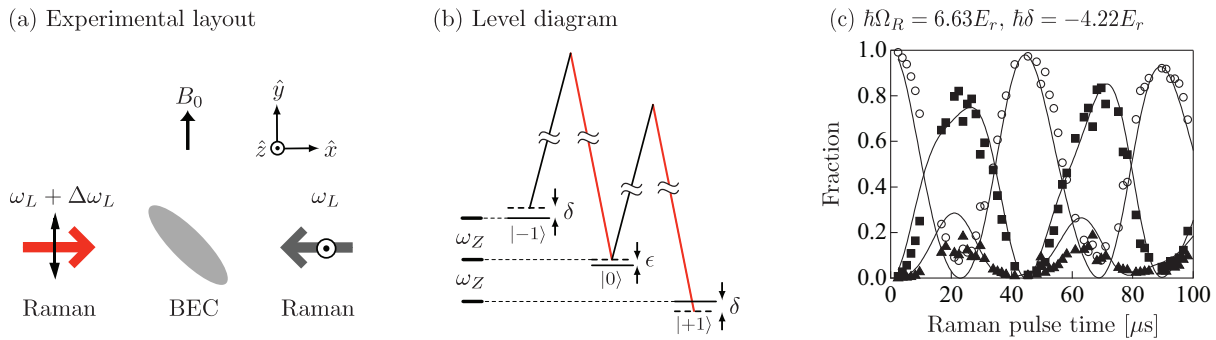}
\end{center}
\caption{(a) The $^{87}$Rb BEC in a dipole trap created by two 1550
nm crossed beams in a bias field $B_{0}\hat{y}$ (gravity is along
$-\hat{z}$). The two Raman laser beams are counter-propagating along
$\hat{x}$, with frequencies $\omega_{L}$ and
$(\omega_{L}+\Delta\omega_{L})$, linearly polarized along $\hat{z}$
and $\hat{y}$, respectively. (b) Level diagram of Raman coupling
within the $F=1$ ground state. The linear and quadratic Zeeman
shifts are $\omega_{Z}$ and $\epsilon$, and $\delta$ is the Raman
detuning. (c) As a function of Raman pulse time, we show the
fraction of atoms in $|m_{F}=-1,k_x=0\rangle$ (open circles),
$|0,-2k_r\rangle$ (solid squares), and $|+1,-4k_r\rangle$ (solid
triangles), the states comprising the $\Psi(\tilde{k}_x=-2k_r)$
family. The atoms start in $|-1,k_{x}=0\rangle$, and are nearly
resonant for the $|-1,0\rangle \rightarrow |0,-2k_r\rangle$
transition at $\hbar \delta =-4.22E_r$. We determine $\hbar
\Omega_{R}= 6.63(4)E_r$ by a global fit (solid lines) to the
populations in $\Psi(-2k_r)$.} \label{ramanrabi}
\end{figure*}

Here we report a first step toward using light-induced vector
potentials to simulate magnetic fields. By dressing a BEC with two
counter-propagating Raman laser beams, we realize a spatially
uniform vector gauge potential. The Raman beams couple internal
(spin) states with linear momenta differing by twice the photon
momentum. This gives rise to a spatial gradient (along the Raman
beam direction) of the phase difference between spin components of
the dressed state. As we will show, this spatially varying state
leads to a non-zero vector potential when the coupling is detuned
from Raman resonance. We adiabatically load the BEC into the dressed
state, and measure properties of the dressed-state dispersion
relation by probing its spin and momentum decomposition. Although
the atoms are stationary in the lab frame (having zero group
velocity), the momenta of the individual spin components composing
the dressed state show a non-zero phase velocity which depends on
the strength and detuning of the Raman coupling. Our measurements
agree with a simple single-particle model, and demonstrate the
presence of an effective vector potential.

\begin{figure}
\begin{center}
\includegraphics[width=3.3in]{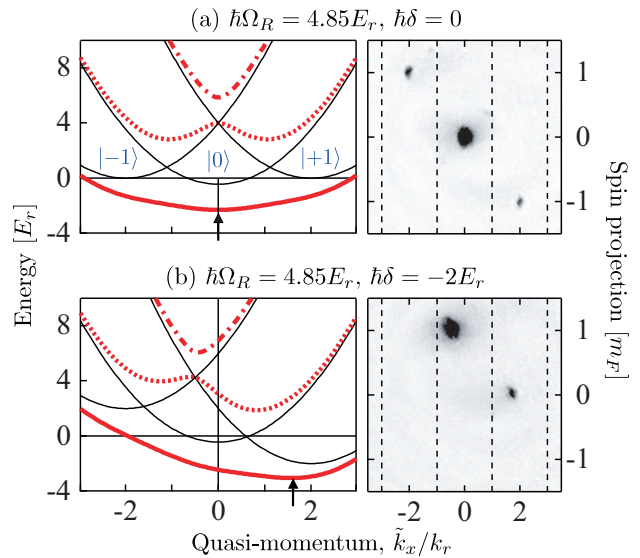}
\end{center}
\caption{Left panels: Energy-momentum dispersion curves
$E(\tilde{k}_x)$ for $\hbar \epsilon=0.44 E_r$ and detuning $\hbar
\delta= 0$ in (a) and $\hbar \delta= -2E_r$ in (b). The thin solid
curves denote the three states
$|-1,\tilde{k}_x+2k_r\rangle,|0,\tilde{k}_x\rangle,|+1,\tilde{k}_x-2k_r\rangle$
in absence of Raman coupling; the thick solid, dotted and
dash-dotted curves indicate dressed states at Raman Rabi frequency
$\Omega_R=4.85 E_r/\hbar$. The arrows indicate
$\tilde{k}_x=\tilde{k}_{\rm min}$ in the $j=1$ dressed state. Right
panels: Time-of-flight images of the Raman-dressed state at
$\hbar\Omega_R=4.85(35) E_r$, for detuning $\hbar \delta=0$ in (a),
and $\hbar\delta=-2 E_r$ in (b). The Raman beams are along
$\hat{x}$, and the three spin and momentum components,
$|-1,\tilde{k}_{\rm min}+2k_r\rangle,|0,\tilde{k}_{\rm min}\rangle$
and $|+1,\tilde{k}_{\rm min}-2k_r\rangle$, are separated along
$\hat{y}$ (after a small shear in the image which realigns the
Stern-Gerlarch gradient direction along $\hat{y}$).}
\label{rabisimu}
\end{figure}

We dress a $^{87}$Rb BEC in the $F$=1 ground state with two Raman
laser beams counter-propagating along $\hat{x}$. Together these
beams couple states $|m_F,k_x\rangle$ differing in internal angular
momentum by $\hbar$ ($\Delta m_F=\pm 1$), and differing in linear
momentum $k_x$ by $2k_r$. Here, $k_r=2 \pi /\lambda$ is the
single-photon recoil momentum, and $\lambda$ is the wavelength of
the Raman beams. We define $E_r=\hbar^{2}k_{r}^{2}/2m$ as the recoil
energy. The family of three states coupled by the Raman field,
$\Psi(\tilde{k}_x)=\left\{|-1,\tilde{k}_x+2k_r\rangle,|0,\tilde{k}_x\rangle,|+1,\tilde{k}_x-2k_r\rangle
\right\}$, are labeled by the quasi-momentum, $\tilde{k}_x$. The
Raman beams have frequencies of $\omega_L$ and $(\omega_{L}+\Delta
\omega_L)$ ($\Delta \omega_{L}>0$), and a bias field $B_0\hat{y}$
produces a Zeeman shift $\hbar \omega_{Z}=g\mu_{B}B_0 \simeq \hbar
\Delta \omega_{L}$ (see Fig.~\ref{ramanrabi}ab). Since the momentum
transfer is only along $\hat{x}$, the single-particle Hamiltonian
can be written as ${\cal H} ={\cal
H}_{1}(k_x)+\left[\hbar^2(k_y^2+k_z^2)/2m+V(\vec{r})\right]\otimes{\bf1}$,
where ${\cal H}_1$ is the Hamiltonian for the Raman coupling, the
Zeeman energies and the motion along $\hat{x}$, and ${\bf1}$ is the
$3\times 3$ unit matrix acting on the spin space. $V(\vec{r})$ is
the state-independent trapping potential (arising from a
far-off-resonance dipole trap and the scalar light shift of the
Raman beams), and $m$ is the atomic mass. In the rotating wave
approximation for the frame rotating at $\Delta \omega_{L}$, ${\cal
H}_{1}/\hbar$ expressed in the state basis of the family
$\Psi(\tilde{k}_x)$ is
\begin{eqnarray}\nonumber \left(
\begin{array}{ccc}{\frac{\hbar}{2m}(\tilde{k}_x+2k_r)^2-\delta}&\Omega_{R}/2&0\\\Omega_{R}/2&\frac{\hbar}{2m}\tilde{k}_{x}^2-\epsilon&\Omega_{R}/2
\\0&\Omega_{R}/2&{\frac{\hbar}{2m}(\tilde{k}_x-2k_r)^2+\delta}\end{array}\right).
\end{eqnarray}
Here $\delta=(\Delta \omega_{L}-\omega_{Z})$ is the detuning from
Raman resonance
%$\epsilon=U_{-1}-U_{0}=U_{0}-U_{1}$, where $U_{m_F}$ is the stark
, $\Omega_R$ is the resonant Raman Rabi frequency, and $\epsilon$
accounts for a small quadratic Zeeman shift (Fig.~\ref{ramanrabi}b).
For each $\tilde{k}_x$, diagonalizing ${\cal H}_1$ gives three
energy eigenvalues $E_{j}(\tilde{k}_x)$ ($j=1,2,3$). For dressed
atoms in state $j$, $E_{j}(\tilde{k}_x)$ is the effective dispersion
relation, which depends on experimental parameters, $\delta$,
$\Omega_{R}$, and $\epsilon$ (left panels of Fig.~\ref{rabisimu}).
For example, the number of energy minima (from one to three) and
their positions, $\tilde{k}_{\rm min}$, are experimentally tunable.
Around each $\tilde{k}_{\rm min}$, the dispersion can be expanded as
$E(\tilde{k}_x)\approx \hbar^2(\tilde{k}_x-\tilde{k}_{\rm
min})^2/2m^*$, where $m^*$ is an effective mass. In this expansion,
we identify $\tilde{k}_{\rm min}$ with the light-induced vector
gauge potential, in analogy to the Hamiltonian for a particle of
charge $q$ in the usual magnetic vector potential $\vec{A}$:
$(\vec{p}-q\vec{A})^2/2m$. In our experiment, we load a trapped BEC
into the lowest energy $j=1$ dressed state, and measure its
quasi-momentum, equal to $\tilde{k}_{\rm min}$ for adiabatic
loading.

Our experiment starts with a 3D $^{87}$Rb BEC in a combined
magnetic-quadrupole plus optical trap, as described in
Ref.~\cite{Lin08}. We transfer the atoms to an all-optical crossed
dipole trap, formed by a pair of 1550~nm beams, by ramping the
quadrupole field gradient to zero. The trapping beams are aligned
along $\hat{x}-\hat{y}$ (horizontal beam) and at $\sim 10^{\circ}$
from $\hat{z}$ (vertical beam). A uniform bias field along $\hat{y}$
gives a linear Zeeman shift $\omega_{Z}/2\pi \simeq 3.25$~MHz and a
quadratic shift $\epsilon/2\pi=1.55$~kHz. The optically-trapped
condensate typically has $N= 2.5 \times 10^5$ atoms in
$|m_{F}=-1,k_x=0\rangle$, with measured trap frequencies of $\approx
30$~Hz parallel to, and $\approx 95$~Hz perpendicular to, the
horizontal beam.

In order to Raman-couple states differing in $m_{F}$ by $\pm 1$, the
$\lambda=804.3$~nm Raman beams are linearly polarized along
$\hat{y}$ and $\hat{z}$, corresponding to $\pi$ and $\sigma$
relative to the quantization axis $\hat{y}$. The beams have $1/e^2$
radii of $180(20)\ \mu$m~\footnote{Uncertainties reflect the
uncorrelated combination of 1-$\sigma$ statistical and systematic
uncertainties.}, much larger than the $20\ \mu$m BEC. These beams
give a total scalar light shift up to $60E_r$, where $E_r=h\times
3.55$~kHz, and contribute an additional harmonic potential with
frequency up to $50$~Hz along $\hat{y}$ and $\hat{z}$. The
differential light shift between adjacent $m_F$ states arising from
the combination of misalignment and imperfect polarization is
estimated to be smaller than $0.2 E_r$. We determine $\Omega_R$ by
observing population oscillations driven by the Raman beams and
fitting to the expected behavior, as shown in Fig.~\ref{ramanrabi}c.

\begin{figure}
\includegraphics[width=3.44in]{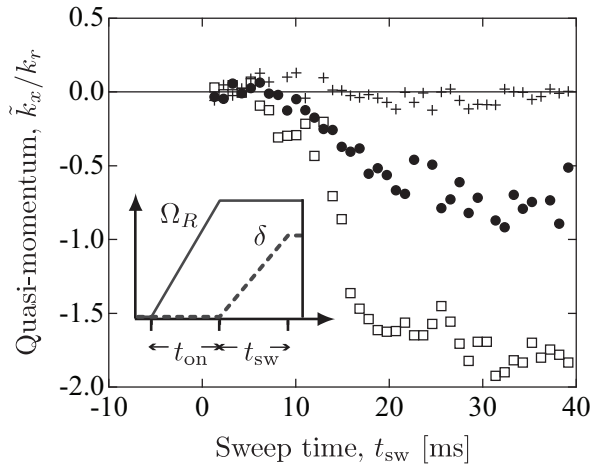}
\caption{Quasi-momentum $\tilde{k}_x$ of the Raman-dressed state
versus sweeping time $t_{\rm sw}$ of Raman detuning from $\hbar
\delta=0$ to $\hbar \delta=1 E_r$ (solid circles) and $\hbar
\delta=2E_r$ (open squares), at $\hbar \Omega_R=5E_r$. The crosses
denote $\tilde{k}_{x}=0$ without sweeping ($\hbar\delta=0$). The
inset shows the time sequence of $\Omega_R$ and $\delta$ for the
loading into a detuned Raman-dressed state. } \label{loading}
\end{figure}

We developed a procedure to adiabatically load the $|-1,k_x=0
\rangle$ BEC into the lowest energy, $j=1$, Raman-dressed state. For
a BEC initially in $|0,k_x=0 \rangle$, this could be achieved simply
by slowly turning on the Raman beams at detuning $\delta=0$,
resulting in the $j=1$, $\tilde{k}_x=0$ dressed state, located at
the minimum of $E_{j=1}(\tilde{k}_x)$. However, our initial state
$|-1,k_x=0 \rangle$, for which $k_x=\tilde{k}_x+2k_r$, is a member
of the family of states $\Psi(\tilde{k}_x=-2k_r)$. This simple
loading procedure would therefore result in the $j=1$,
$\tilde{k}_x=-2k_r$ dressed state. In order to transfer our
$\tilde{k}_x=-2k_r$ BEC to the desired $\tilde{k}_x=0$ dressed
state, we use an additional rf coupling which mixes families of
states with $\tilde{k}_x$ differing by $2k_r$. The rf frequency is
always equal to the frequency difference between two Raman beams,
$\Delta \omega_{L}/2\pi= 3.250$~MHz. The loading sequence is as
follows: (i) We turn on the rf coupling in 1 ms to a resonant Rabi
frequency $\Omega_{\rm rf}/2\pi$= 12 kHz, with an initial detuning
$\hbar \delta=15 E_r$. We then sweep the detuning to resonance
\footnote{All the detuning ramps are linear versus time. The Raman
beam power, swept by applying a linear ramp to the AOM controlling
its power, turns on nearly quadratically.} in 9 ms ($\delta$ is
varied by ramping the bias field $B_0$, leaving $\Delta \omega_{L}$
constant.), loading the atoms into the lowest energy rf-dressed
state, a spin superposition still at $k_x=0$. (ii) We ramp on the
Raman coupling in $t_{\rm on}=$ 20 ms to a variable Rabi frequency
$\Omega_R$, and then turn off the rf in 2 ms. This loads the BEC
into the $j=1$, $\tilde{k}_x=0$ dressed state. In the dressed state,
the BEC heats at $\sim 0.3$~nK/ms, we believe due to technical noise
in $B_0$ or $\Delta \omega_{L}$. In addition, we observe unwanted
population in the $j=2$ dressed state. Therefore, as part of step
(ii), we further evaporatively cool by decreasing the intensity of
the horizontal trapping beam by $20\%$ in 20 ms, after which it
remains constant. The evaporation typically reduces the BEC number
to $7 \times 10^4$.

In the final step (iii), we transfer the $\tilde{k}_{x}=0$ dressed
state into a non-zero $\tilde{k}_{x}$ state by sweeping the detuning
from resonance to a variable $\delta$ in $t_{\rm sw}$ and holding
for $t_{\rm h}$. As we show below, for $t_{\rm sw}\gtrsim 20$~ms,
this process adiabatically loads the atoms into the $j=1$
Raman-dressed state at $\tilde{k}_x=\tilde{k}_{\rm min}(\delta)$.

We characterize the Raman-dressed state by abruptly turning off the
dipole trap and the Raman beams in less than 1~$\mu$s, projecting
the atomic state onto its individual spin and momentum components.
The atoms then expand in a magnetic field gradient applied during
time-of-flight (TOF) approximately along $\hat{y}$, and the three
spin states spatially separate due to the Stern-Gerlach effect.
Imaging the atoms after a 20~ms TOF gives the momentum and spin
composition of the dressed state (see right panels of
Fig.~\ref{rabisimu}). The quasi-momentum $\tilde{k}_x$ of the BEC is
given by the shift along $\hat{x}$ of all three spin components from
their positions when $\tilde{k}_{x}=0$. We determine
$\tilde{k}_{x}=0$ to within 0.03~$k_r$ by using the loading
procedure above, except that in (ii), the Raman frequencies are
sufficiently far-detuned that the atoms experience the scalar light
shift but are not Raman-coupled; then the rf and Raman fields are
snapped off concurrently before TOF.

\begin{figure}
\begin{center}
\includegraphics[width=3.44in]{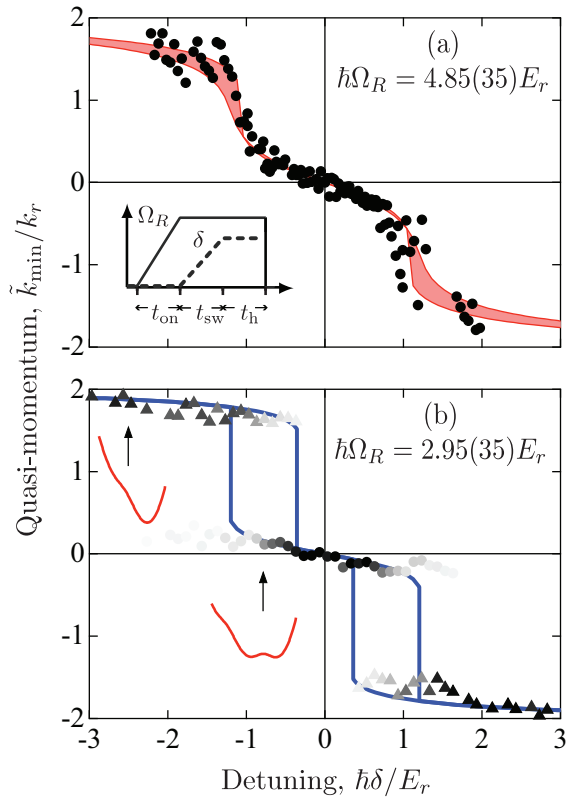}
\end{center}
\caption{(a) Measured quasi-momentum $\tilde{k}_{\rm min}$ versus
detuning $\delta$, and the calculated $\tilde{k}_{\rm min}$ at Rabi
frequency $\Omega_R=4.85(35) E_{r}/\hbar$ (with the uncertainty
indicated by the shaded region). The inset shows the time sequence
of $\Omega_R$ and $\delta$. (b) Quasi-momentum $\tilde{k}_{\rm
min1}$ (circles) and $\tilde{k}_{\rm min2}$ (triangles) versus
detuning $\delta$, and the calculated $\tilde{k}_{\rm min}$ (solid
lines) at $\hbar \Omega_{R}=2.95(35)E_{r}$. The intensity of the
circles indicates the population of $|0,\tilde{k}_{\rm min1}
\rangle$ relative to its value at $\delta=0$. The intensity of the
triangles for $\delta<0$ ($\delta>0$) indicates the population of
$|+1,\tilde{k}_{\rm min2}-2k_r \rangle$ ($|-1,\tilde{k}_{\rm
min2}+2k_r \rangle$) relative to its value at $\hbar\delta=-3 E_r$
($\hbar\delta=3E_r$). The insets picture $E(\tilde{k_x})$ at
$\hbar\delta=-0.8$ and $-2.5 E_r$ as indicated by the
arrows.}\label{largesmallrabi}
\end{figure}

It is straightforward to adiabatically load into the rf-dressed
state in step (i). To ensure the subsequent adiabatic loading from
the rf-dressed state into the Raman-dressed state in step (ii), we
ramped up the Raman power in a time $t_{\rm on}$, then ramped it off
in $t_{\rm on}$. For $t_{\rm on} \geq 2$~ms, there is no discernable
excitation into other rf-dressed states. We conservatively use
$t_{\rm on}=20$~ms as the loading time in step (ii). To determine
$t_{\rm sw}$ in step (iii), we measure $\tilde{k}_x$ as a function
of $t_{\rm sw}$ for negligible $t_{\rm h}$. Figure~\ref{loading}
shows $\tilde{k}_x$ versus $t_{\rm sw}$ at $\hbar \delta=0,1,2E_r$
and $\hbar \Omega_R=5 E_r$, for $t_{\rm h}=0.1$~ms. We use $t_{\rm
sw}=20$~ms, the time at which $\tilde{k}_{x}$ has nearly reached its
equilibrium value. This adiabatic following is enabled by the
external trap, and the time scale is comparable to our typical
trapping periods. After $t_{\rm sw}$, we add a hold time $t_{\rm
h}=20$~ms during which residual excitations damp.

Figure~\ref{rabisimu} shows spin-resolved TOF images of
adiabatically loaded Raman-dressed states at $\hbar
\Omega_R=4.85(35) E_r$ for $\hbar \delta=-2$ and $0 E_r$. The
resonance condition, $\delta=0$, is determined from the symmetry of
the rf-dressed state \footnote{On resonance the rf-dressed state has
equal population in $|-1,k_{x}=0\rangle$ and $|+1,k_{x}=0\rangle$,
even in the presence of a quadratic Zeeman shift. The resonance
condition $\delta$=0 determined from the symmetry of the Raman
dressing is within 0.25$E_r$ of that from the rf-dressed states.},
with an uncertainty of $h \times1.5$~kHz=0.4$E_r$, limited by the
stability of $B_0$. The quasi-momentum $\tilde{k}_{x}=\tilde{k}_{\rm
min}$, measured as a function of $\delta$, is shown in
Fig.~\ref{largesmallrabi}a, along with the calculated
$\tilde{k}_{\rm min}(\delta)$. Each different $\tilde{k}_{\rm min}$
represents a spatially uniform vector gauge potential, analogous to
the magnetic vector potential, with $\tilde{k}_{\rm
min}=q\vec{A}/\hbar$. For such a uniform $\vec{A}$, the magnetic
field $\vec{B}=\nabla \times \vec{A}$ would be zero.

%\begin{figure}
%\includegraphics[width=3.0in]{smallrabi2}
%\caption{Quasi-momentum $\tilde{k}_{\rm min1}$ (circles) and
%$\tilde{k}_{\rm min2}$ (triangles) versus detuning $\delta$ at Rabi
%frequency $\Omega_{R}=2.95(35)E_{r}/\hbar$. The intensity of the
%symbols indicates the relative population of $|0,\tilde{k}_{\rm
%min1} \rangle$ (circles), $|+1,\tilde{k}_{\rm min2}-2k_r \rangle$
%(triangles) at $\delta<0$, and $|-1,\tilde{k}_{\rm min2}+2k_r
%\rangle$ (triangles) at $\delta>0$. The solid lines are the
%calculation for $\hbar \Omega_R=2.95 E_r$. The inset shows the TOF
%image at $\hbar \delta= -0.8 E_r$, and energy-momentum dispersion
%curves $E_{j=1}(\tilde{k}_x)$ at $\hbar \delta=-0.8,-2.5
%E_r$.}\label{smallrabi}
%\end{figure}

For Rabi frequency $\Omega_R \leq 4.47 E_r/\hbar$,
$E_{j=1}(\tilde{k}_x)$ has multiple minima for some detunings
$\delta$. For data taken as in Fig.~\ref{largesmallrabi}a,
Fig.~\ref{largesmallrabi}b shows such multiple $\tilde{k}_{\rm
min}$, expected for $\delta_{a}<|\delta|<\delta_{b}$, where $\hbar
\delta_{a}=0.35(20) E_r$ and $\hbar \delta_{b}=1.2(1)E_r$, at $\hbar
\Omega_R=2.95(35) E_r$. At small and large detuning, we observe only
one $\tilde{k}_{\rm min}$. However, at intermediate detuning, we
observe two $\tilde{k}_{\rm min}$ at $\tilde{k}_{\rm min1}$ and
$\tilde{k}_{\rm min2}$~\cite{stanescu08}. In the figure, the
relative population is indicated by the gray scale, and the solid
line is the calculated value(s). The observed values of
$\tilde{k}_{\rm min}(\delta)$ agree well with the theory, and a
simultaneous fit of $\tilde{k}_{\rm min1}$ and $\tilde{k}_{\rm
min2}$ with $\Omega_R$ as a free parameter gives $\hbar
\Omega_R=2.9(3) E_r$, which agrees well with the independently
measured $\hbar \Omega_R=2.95(35) E_r$. As expected, we observe
population in only one $\tilde{k}_{x}$ for $|\delta|<\delta_{a}$,
however, for some $|\delta|>\delta_{b}$, there is discernable
population in two $\tilde{k}_{x}$ where we expect only one
$\tilde{k}_{\rm min}$. We attribute this to non-adiabicity when the
minimum at $\tilde{k}_{\rm min1}$ disappears during the detuning
sweep in loading step (iii).

%\footnote{We determine the differential stark shift $\epsilon$ from
%$\delta \tilde{k}_x(\delta)=0$ at $\delta=\epsilon$, which are
%$\epsilon=-0.02\pm 0.4$ $E_r$ and $\epsilon=0.2\pm0.4$ $E_r$, for
%$\Omega_R=4.85(35)$ $E_r$ and $2.95(35)$ $E_r$, respectively. The
%uncertainty of $\epsilon$ is dominated by that of the bias field,
%$\sigma_{f_0}$, therefore $\epsilon$ is comparable to or smaller
%than $\sigma_{f_0}$. This is consistent with the measured
%ellipticity of the polarizations of the Raman beams, smaller than
%$1\%$, which gives the upper bound of $\epsilon=$ 0.15 $E_r$, at
%$\lambda=804.3$ nm and $\Omega_R=4.8 E_r$.}.

In conclusion, we have prepared a Bose-Einstein condensate in a
Raman-dressed state with a non-zero quasi-momentum, controlled by
the Rabi frequency $\Omega_R$ and the detuning $\delta$. This
technique, in conjunction with a spatial gradient of the bias field
(therefore the detuning) along $\hat{y}$, gives a spatial gradient
of light-induced momentum $\tilde{k}_{\rm min}(y)\hat{x}$ and
creates an effective magnetic field along $\hat{z}$
\cite{Spielman08}. The analog of the magnetic length
$l_B=\sqrt{\hbar/(qB)}$ (the classical cyclotron radius of an orbit
with one unit of angular momentum $\hbar$) is
$(\partial{\tilde{k}_{\rm min}}/\partial y)^{-1/2}$, and is about
the spacing between vortices in a vortex lattice like that formed in
a slowly rotating BEC~\cite{schweikhard04}. For a gradient in
$\tilde{k}_{\rm min}$ of one $k_r$ across a condensate of radius
$R$, this gives an analog magnetic length $\approx$ 1.4~$\mu$m for a
$R\approx10\ \mu m$ 3D BEC, and we expect a $\sim 25$-vortex lattice
to form in the condensate. For dilute 2D systems, this is sufficient
to reach the quantum Hall regime.

This work was partially supported by ONR, ODNI, and ARO with funds
from the DARPA OLE program. R.L.C. acknowledges the NIST/NRC
postdoctoral program.

%\bibliographystyle{pra}

%\bibliography{ylref3}

\end{document}